\documentclass[conference]{IEEEtran}
\IEEEoverridecommandlockouts
\makeatletter
\newcommand{\linebreakand}{%
\end{@IEEEauthorhalign}
\hfill\mbox{}\par
\mbox{}\hfill\begin{@IEEEauthorhalign}
}
\usepackage{cite}
\usepackage{amsmath,amssymb,amsfonts}
\usepackage{algorithm}
\usepackage{algpseudocode}
\usepackage{subfig}
\usepackage[acronym]{glossaries}
\usepackage{optidef}
\usepackage{multirow} 
\usepackage{hyperref}
\usepackage{xcolor}
\usepackage{fancyhdr}
	
\DeclareMathAlphabet{\mathbbmsl}{U}{bbm}{m}{sl}
\newtheorem{proof}{Proof}

\newtheorem{lemma}{Lemma}

\newacronym{bs}{BS}{base station}
\newacronym{eh}{EH}{energy harvesting}
\newacronym{rf}{RF}{radio frequency}
\newacronym{tsr}{TSR}{time switching ratio}
\newacronym{ts}{TS}{time switching}
\newacronym{uav}{UAV}{unmanned aerial vehicle}
\newacronym{ur}{UR}{\gls{uav}-enabled relaying}
\newacronym{iot}{IoT}{Internet of Things}
\newacronym{Id}{ID}{IoT device}
\newacronym{Ed}{ED}{edge device}
\newacronym{bis}{BIS}{best IDs selection}
\newacronym{bfd}{BFD}{best far device}
\newacronym{bnd}{BND}{best near device}
\newacronym{pcsi}{pCSI}{perfect channel state information}
\newacronym{icsi}{iCSI}{imperfect channel state information}
\newacronym{psic}{pSIC}{perfect SIC}
\newacronym{isic}{iSIC}{imperfect successive interference cancellation}
\newacronym{sic}{SIC}{successive interference cancellation}
\newacronym{noma}{NOMA}{nonorthogonal multiple access}
\newacronym{af}{AF}{amplify-and-forward}
\newacronym{df}{DF}{decode-and-forward}
\newacronym{cdf}{CDF}{cumulative distribution function}
\newacronym{pdf}{PDF}{probability density function}
\newacronym{sinr}{SINR}{signal-to-interference-plus-noise ratio}
\newacronym{snr}{SNR}{signal-to-noise ratio}
\newacronym{awgn}{AWGN}{additive white Gaussian noise}
\newacronym{rv}{RV}{random variable}
\newacronym{oma}{OMA}{orthogonal multiple access}
\newacronym{los}{LoS}{Line-of-Sight}
\newacronym{nos}{NLoS}{non-LoS}
\newacronym{mec}{MEC}{mobile-edge computing}
\newacronym{an}{AN}{artificial noise}
\newacronym{pls}{PLS}{physical-layer security}
\newacronym{wpt}{WPT}{wireless power transfer}
\newacronym{op}{OP}{outage probability}
\newacronym{scp}{SCP}{successful computation probability}
\newacronym{pso}{PSO}{particle swarm optimization}
\newacronym{rcga}{RCGA}{real-coded genetic algorithm}
\newacronym{re}{RF EH}{\gls{rf} \gls{eh}}
\newacronym{sscp}{SSCP}{secrecy successful computation probability}
\newacronym{bish}{BIH}{BIS of cluster HP}
\newacronym{bisl}{BIL}{BIS of cluster LP}

\def\BibTeX{{\rm B\kern-.05em{\sc i\kern-.025em b}\kern-.08em
    T\kern-.1667em\lower.7ex\hbox{E}\kern-.125emX}}

\begin{document}
\title{Secrecy Offloading Analysis of UAV-assisted NOMA-MEC Incorporating WPT in IoT Networks}
\author{
	\IEEEauthorblockN{Gia-Huy Nguyen}
	\IEEEauthorblockA{\textit{ICT Department} \\
		\textit{FPT University}\\
		Hanoi 10000, Vietnam \\
		huynghe180064@fpt.edu.vn}
\and
	\IEEEauthorblockN{Anh-Nhat Nguyen}
	\IEEEauthorblockA{\textit{ICT Department} \\
		\textit{FPT University}\\
		Hanoi 10000, Vietnam \\
		nhatna3@fe.edu.vn}
\and
\IEEEauthorblockN{Minh-Sang Nguyen}
\IEEEauthorblockA{\textit{ICT Department} \\
	\textit{FPT University}\\
	Hanoi 10000, Vietnam \\
	sangnmhe176048@fpt.edu.vn}
\and
\IEEEauthorblockN{Khai Nguyen}
\IEEEauthorblockA{\textit{ICT Department} \\
	\textit{FPT University}\\
	Hanoi 10000, Vietnam \\
	khainhe176049@fpt.edu.vn}
\linebreakand
\IEEEauthorblockN{Tung-Son Ngo}
\IEEEauthorblockA{\textit{ICT Department} \\
	\textit{FPT University}\\
	Hanoi 10000, Vietnam \\
	sonnt69@fe.edu.vn}
\and 
\IEEEauthorblockN{Ngoc-Anh Bui}
\IEEEauthorblockA{\textit{ICT Department} \\
	\textit{FPT University}\\
	Hanoi 10000, Vietnam \\
	anhbn5@fe.edu.vn}
\and
\IEEEauthorblockN{Phuong-Chi Le}
\IEEEauthorblockA{\textit{ICT Department} \\
	\textit{FPT University}\\
	Hanoi 10000, Vietnam \\
	chilp2@fe.edu.vn}
\and
\IEEEauthorblockN{Manh-Duc Hoang}
\IEEEauthorblockA{\textit{ICT Department} \\
	\textit{FPT University}\\
	Hanoi 10000, Vietnam \\
	duchm29@fe.edu.vn}
}
	
\maketitle
\thispagestyle{fancy}

\begin{abstract}
This article studies the efficiency of secrecy data offloading for an \gls{uav}-assisted \gls{noma}-integrated \gls{mec} incorporating \gls{wpt} within an \gls{iot} network. Specifically, this study assumes an \gls{uav} to function in dual roles: as a mobile computation platform and as an aerial power-supply station, offering substantial advantages for resource-constrained \glspl{Ed} in mitigating interference from an passive eavesdropper. To assess the system's secrecy offloading efficacy, the \gls{sscp} closed-formed formulation under Nakagami-$m$ fading channel is derived. The theoretical results are conducted with a variety of parameters, thereby validating the precision of our analysis. 
\end{abstract}


\section{Introduction}\label{sec:in}
The advent of next generation networks has significantly enhanced connectivity and data interchange propagation, paving the way for the rapid expansion of \acrfull{iot} devices \cite{ZZh,JCo,NA22}. While these networks are acquired with the capability of delivering high-speed data transfer and extensive coverage, they also introduce inevitable deficiencies, such as network congestion and security vulnerabilities \cite{PSh,SAh}.

\Acrfull{noma} and \gls{pls} have offered viable solutions to tackle the incoming challenges \cite{NWa}. By utilizing the power domains, \gls{noma} facilitates the concurrent data transmission across a multitude of mobile users within identical time resources, thereby promoting the spectral efficiency and user fairness \cite{RZh}. On the other hand, by exploiting the physical properties of the wireless communication, \gls{pls} is capable of safeguarding data confidentiality against hostile attacks or eavesdroppers, thus enhancing the secrecy of \gls{noma}-applied systems \cite{NA24}. However, the high-speed \gls{noma} services cause substantial energy consumption in mobile devices, which directly contributes to the computation-constrained issue. 

Thus, \acrfull{wpt} and \acrfull{mec} have emerged as pivotal technologies \cite{NA22_2}. \gls{wpt} offers a promising solution to the energy demands of \gls{iot} networks by enabling the continuous wireless power-supply to mobile devices, reducing the need of battery replacements \cite{HNG}. Furthermore, to address the data computation issue, the key feature of \gls{mec} permits the mobile devices to offload computation-intensive tasks to the edge severs \cite{XZh}. Hence, the network robustness can be enhanced \cite{ISB}.

In addition, current research has increasingly focused on the evaluations of the \acrfull{uav} in \gls{noma}-\gls{mec} systems \cite{NA23}. The unique features of \gls{uav}, including flexible deployments, the establishment of strong \gls{los} transmissions, extensive coverage over large areas, contribute to the advantages of providing superior communication quality compared to the conventional installations of terrestrial cellular networks.

Driven by the preceding discourses, this study examines an \gls{uav}-assisted  \gls{noma}-\gls{mec} with \gls{wpt} system under Nakagami-$m$ fading channels for \gls{iot} networks, with the interference of an passive eavesdropper. Additionally, the study enhances the communication interchanges between \gls{uav} and \acrfullpl{Ed} by employing the probabilistic models of \gls{los} and \gls{nos} propagation. The primary contributions of our paper are as follows:
\begin{itemize}
	\item We investigate the secrecy offloading efficiency for an \gls{uav}-assisted \gls{noma}-integrated \gls{mec} processor incorporating \gls{wpt} within an \gls{iot} network. Hence, we introduce a system protocol that ensures an efficient wireless charging and offloading operation. 
	
	\item The \acrfull{sscp} closed-formed formulation of the whole system in regards of \gls{icsi} and \gls{isic} is derived. 
	
	
	\item Theoretical findings are conducted with a variety of parameters including \gls{uav}'s transmission power, the number of \glspl{Ed}, network parameters, \gls{eh} ratio, and the \gls{uav}'s positions and altitudes, thereby validating the precision of our analysis.
\end{itemize}	

The paper's remaining sections are structured thusly as. Section \ref{sec:sm-cm}  provides insights of the system model's characteristics and its protocol. Thereafter, Section~\ref{sec:pa} delves into the derivation of the system metric formulation. Section~\ref{sec:nr} discusses the theoretical findings. Eventually, 	Section~\ref{sec:conc} presents the conclusions of this study.
\section{System Model and Protocol}\label{sec:sm-cm}
\subsection{System and Channel Model}
\begin{figure}[!]
	\centering{\includegraphics[width=0.5\textwidth]{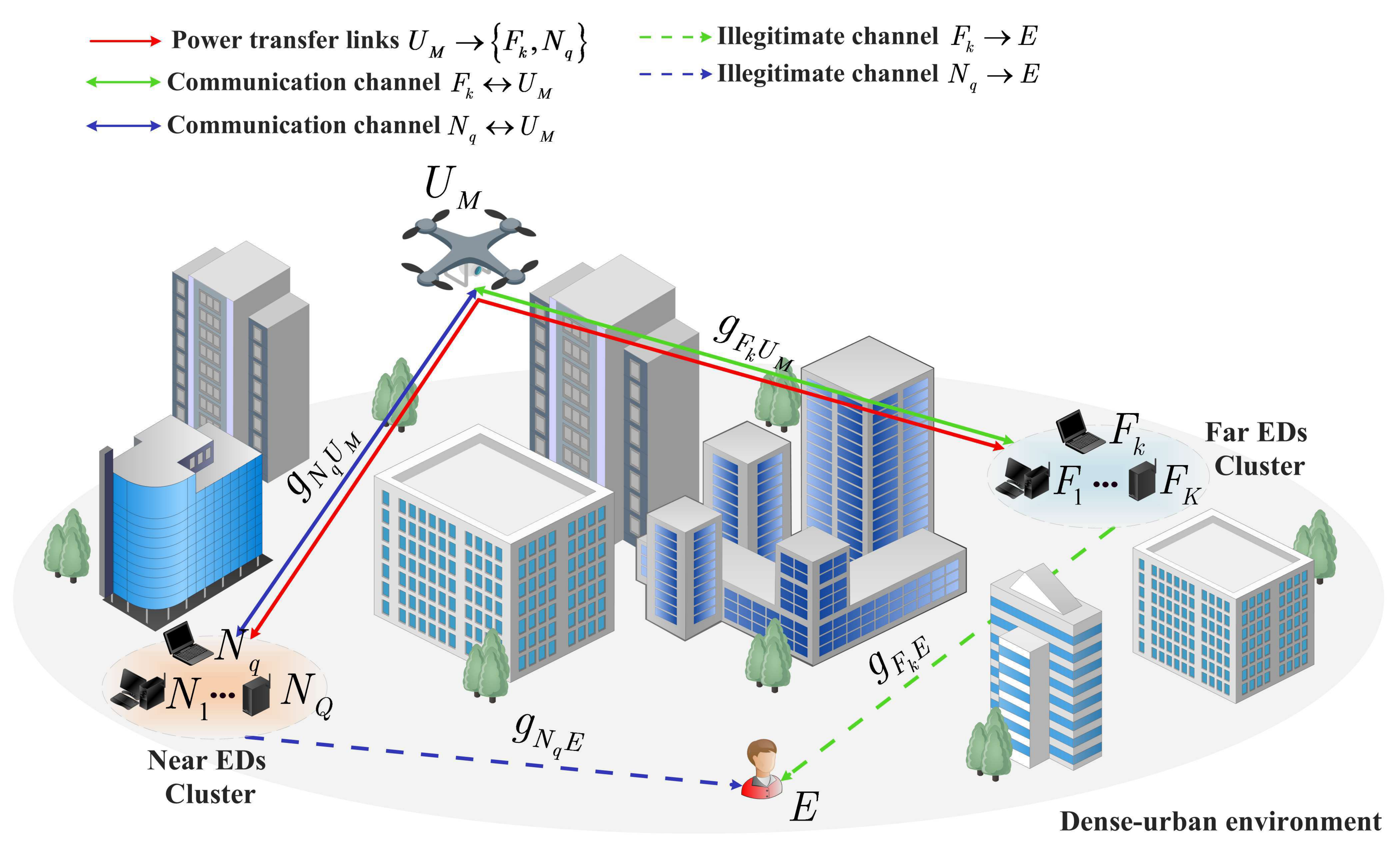}}
	\caption{
		System model of the \gls{uav}-assisted uplink \gls{noma}-\gls{mec} with \gls{wpt} in \gls{iot} networks.
	}
	\label{fig:sm}
\end{figure}
Fig.~\ref{fig:sm} illustrates a system model of a dense-urban scenario in the \gls{iot} networks comprising of an \gls{uav}-assisted \gls{noma}-integrated \gls{mec} processor, denoted by $U_M$, a remote cluster of \glspl{Ed} $\left({{{F}}_{k}}\right)$ with $K$ devices, and a nearby \glspl{Ed} cluster $\left({N}_q\right)$ with $Q$ devices, where $k\in \left\{ 1,\text{ }...,K \right\}$ and $q\in \left\{ 1,\text{ }...,Q \right\}$. Particularly, $U_M$ initiates the system operations by broadcasting the wireless power via radio frequency signals to the two edge clusters. These signals are then harvested by those power-constrained \glspl{Ed}. Following the \gls{eh} phase, a selection process is employed to identify the \gls{Ed} with the highest computational efficiency from each cluster, which is utilized to perform secure data offloading tasks to the $U_M$ by uplink \gls{noma}, in the interference of an passive eavesdropper $\left(E\right)$. All devices are assumed to be equipped with a single antenna and operate in a half-duplex mode. 

Furthermore, it is posited that each \gls{Ed} carries out identical workloads with length $l$ (bits), and these tasks are independently separated into distinctive groups. Thus, the offloading capacities of the \glspl{Ed} are expressed as \cite{HNG} $\mathcal{C}_{\vartheta}^{off}={{\sigma }_{\vartheta}}l$, where ${{\sigma }_{\vartheta }}$ represents the offloading ratio, with $\left( 0<{{\sigma }_{\vartheta }}<1 \right)$, and $\vartheta\in\left(F_k, N_q\right)$.

To model spatial relationships of the system components, we exploit a 3D Cartesian coordinate system, in which the \gls{uav}'s coordinate is ${{U}_{M}}\left( {{x}_{U}},{{y}_{U}},{{h}_{U}} \right)$, while the ground nodes are denoted as ${{{F}}_{k}}\left( {{x}_{k}},{{y}_{k}},0 \right)$, ${{{N}}_{q}}\left( {{x}_{q}},{{y}_{q}},0 \right)$, and ${{{E}}}\left( {{x}_{E}},{{y}_{E}},0 \right)$. Assuming that the large-scale fading channel between \gls{uav} and \glspl{Ed} is characterized by the probabilistic model of \gls{los} and \gls{nos} propagation. Thus, the mean path-loss is expressed as follows \cite{RZh}:
\begin{equation} \label{mlos}
	{{\overline{\mathbbmsl{L}}}_{ab}}=\left[ {{\mathbbmsl{K}}^{NLoS}}+\frac{{{\mathbbmsl{K}}^{LoS}}-{{\mathbbmsl{K}}^{NLoS}}}{1\text{ }+\text{ }{{\tau }_{1}}{{e}^{\left( -\frac{180}{\pi }{{\tau }_{2}}{{\varphi }_{ab}}+{{\tau }_{2}}{{\tau }_{1}} \right)}}} \right]\mathbbmsl{D}_{ab}^{\theta },
\end{equation}
where $ab\in \left\{ {{{F}}_{k}}{{U}_{M}},{{{N}}_{q}}{U}_{M},{{{F}}_{k}}{{E}},{{{N}}_{q}}{{E}} \right\}$, ${{\mathbbmsl{K}}^{LoS}}={{\mu }^{LoS}}{{\left( \frac{c}{4\pi {{f}_{c}}} \right)}^{-1}}$, ${{\mathbbmsl{K}}^{NLoS}}={{\mu }^{NLoS}}{{\left( \frac{c}{4\pi {{f}_{c}}} \right)}^{-1}}$, $c$ and $f_c$ are the speed of light and the carrier frequency, respectively; ${{\mu }^{LoS}}$ and ${{\mu }^{NLoS}}$ are the excessive path-losses of \gls{los} and \gls{nos} connections; $\tau_1$ and $\tau_2$ are the parameters that vary depending on the surrounding environments; ${{\varphi }_{ab}}=\arcsin (\frac{{{h}_{U}}}{{{\mathbbmsl{D}}_{ab}}})$ is the elevation angle; ${{\mathbbmsl{D}}_{ab}}=\sqrt{{{({{x}_{b}}-{{x}_{a}})}^{2}}+{{({{y}_{b}}-{{y}_{a}})}^{2}}+{{({{h}_{b}}-{{h}_{a}})}^{2}}}$ is the distance of $ab$; $\theta$ is the path-loss exponent. 

It is assumed that the channel coefficients, denoted as $g_{ab}$, are independent, and conform the Nakagami-$m$ fading model. Nevertheless, due to the feedback latencies, the system model is inevitably affected by the \gls{icsi} \cite{HNG}. Thus, the channel coefficient is calculated as: ${{g}_{ab}}\text{ }=\text{ }{{\tilde{g}}_{ab}}\text{ }+\text{ }{{\lambda }_{ab}}$, where $\tilde{{g}}_{ab}$ is the estimated channel coefficient, ${\lambda }_{ab}$ is the channel quality parameter, ${{\lambda }_{ab}}\sim \mathcal{C}\mathcal{N}\left( 0,{{\it\Omega }_{ab}} \right)$. Within this paper, the channel quality variance $\it\Omega_{ab}$ is considered to be constant \cite{KDo}. 

Given that the all channel coefficient are affected by the Nakagami-$m$ fading model,
the \gls{pdf} and \gls{cdf} of the channel coefficients ${{\left|	 {{{\tilde{g}}}_{{{\vartheta}^{*}}E}} \right|}^{2}}$ detected at $E$ is expressed as: 
\begin{align} \label{pdf-E}
	{{f}_{{{\left| {{{\tilde{g}}}_{{{\vartheta }^{*}}E}} \right|}^{2}}}}\left( u \right)&=\frac{{{u}^{m-1}}}{\left( m-1 \right)!}{{\left( \frac{m}{{{\xi }_{{{\vartheta }^{*}}E}}} \right)}^{m}}{{e}^{-\frac{mu}{{{\xi }_{{{\vartheta }^{*}}E}}}}}, \\ \label{cdf-E}
	{{F}_{{{\left| {{{\tilde{g}}}_{{{\vartheta }^{*}}E}} \right|}^{2}}}}\left( u \right)&=1-{{e}^{-\frac{mu}{{{\xi }_{{{\vartheta }^{*}}E}}}}}\sum\limits_{s=0}^{m-1}{\frac{1}{s!}{{\left( \frac{mu}{{{\xi }_{{{\vartheta }^{*}}E}}} \right)}^{s}}}.
\end{align}

Accordingly, by calculating the maximal \glspl{snr} among $U_M$, $F_k$, and $N_q$ connections, the channel gains for the two best \glspl{Ed} in the respective clusters are represented as: ${{\left| {{{\tilde{g}}}_{{{\vartheta }^{*}}U}} \right|}^{2}}=\max \left. \left\{ {{\left| {{{\tilde{g}}}_{{{\vartheta}}{{U}_{M}}}} \right|}^{2}} \right. \right\}$, where ${{\vartheta }^{*}}\in \left( {{F}^{*}},{{N}^{*}} \right)$. Thus, the \gls{pdf} and \gls{cdf} of the estimated channel coefficients  ${{\left|	 {{{\tilde{g}}}_{{{\vartheta}^{*}}U}} \right|}^{2}}$ detected at $U_M$ can be expressed as \cite{NA23}: 
\begin{align} \label{pdf-U}
	{{f}_{{{\left| {{{\tilde{g}}}_{{{\vartheta }^{*}}U}} \right|}^{2}}}}\left( u \right)&=\ddot{\mathcal{Z}}\sum\limits_{p}{\left( \mathcal{Z}-1 \right){{u}^{m-1+\bar{p}}}{{e}^{-\frac{um}{{{\xi }_{{{\vartheta }^{*}U}}}}\left( p+1 \right)}}}, \\ \label{cdf-U}
	{{F}_{{{\left| {{{\tilde{g}}}_{{{\vartheta }^{*}}U}} \right|}^{2}}}}\left( u \right)&=\sum\limits_{p}{\left( \mathcal{Z}  \right){{u}^{{\bar{p}}}}{{e}^{-\frac{pum}{{{\xi }_{{{\vartheta }^{*}U}}}}}}}, 
\end{align}
where $\mathcal{Z}\in \left\{ K,Q \right\}$, $p\in \left\{h,t\right\}$, $\sum\limits_{p}{\left( \mathcal{Z}-1 \right)}=\sum\limits_{p=0}^{\mathcal{Z}-1}{{{\Xi }_{p}}{{\left( -1 \right)}^{p}}{{\Phi }_{1,p}}{{\Phi }_{2,p}}}$,  $\Xi_{p}{=\sum\limits_{{{p}_{1}}=0}^{p}{\sum\limits_{{{p}_{2}}=0}^{p-{{p}_{1}}}{...\sum\limits_{{{p}_{m-1}}=0}^{p-...-{{p}_{m-2}}}}}}$, ${{\Phi }_{1,p}}=\left( \begin{matrix} \mathcal{Z} - 1 \\ p \\ \end{matrix} \right)\left( \begin{matrix} p \\ {{p}_{1}} \\ \end{matrix} \right)\left( \begin{matrix} p-{{p}_{1}} \\ {{p}_{2}} \\ \end{matrix} \right)...\left( \begin{matrix} p-...-{{p}_{m-2}} \\ {{p}_{m-1}} \\ \end{matrix} \right)$, ${{\Phi }_{2,p}}={{\prod\limits_{s=0}^{m-2}{\left[ \frac{1}{s!}{{\left( \frac{m}{{{\xi }_{{{\vartheta }^{*}}U}}} \right)}^{s}} \right]}}^{{{p}_{s+1}}}}$ ${{\left[ \frac{1}{\left( m-1 \right)!}{{\left( \frac{m}{{{\xi }_{{{\vartheta }^{*}}U}}} \right)}^{m-1}} \right]}^{p-...-{{p}_{m-1}}}}$, $\ddot{\mathcal{Z}}=\frac{\mathcal{Z}}{\left( m-1 \right)!}{{\left( \frac{m}{{{\xi }_{{{\vartheta }^{*}}U}}} \right)}^{m}}$, $\bar{p}=\left( m-1 \right)\left( p-{{p}_{1}} \right)-\left( m-2 \right){{p}_{2}}-...-{{p}_{m-1}}$, $m\in \left( 2,3,4,... \right)$.

\subsection{System Protocol}
\begin{figure}[!]
	\centering{\includegraphics[width=0.5\textwidth]{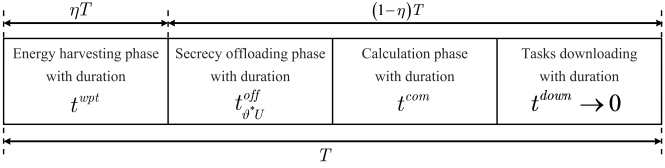}}
	\caption{
		The protocol flowchart for the \gls{uav}-assisted \gls{noma}-\gls{mec} with \gls{wpt} system.
	}
	\label{fig:pr}
\end{figure}
This subsection introduces an quad-phased protocol for the proposed system. Specifically, the protocol flowchart in Fig.~\ref{fig:pr} is described as follows:
\begin{itemize}
	\item In the energy harvesting phase, denoted as $t^{wpt}$: $U_M$ initially sends the radio frequency signals to the \gls{Ed} clusters to harvest energy. Subsequently, the two best \glspl{Ed} from their respective clusters are chosen based on the highest \glspl{snr}. Therefore, the energy conveyed to the \glspl{Ed} is formulated as: ${E_{{{\vartheta }^{*}}U}}=\frac{\beta {{P}_{U}}{{t}^{wpt}} {{\left| {{{g}}}_{{{\vartheta }^{*}}U} \right|}^{2}}}{{{\overline{\mathbbmsl{L}}}_{{{\vartheta }^{*}}U}}}$, where $\beta$ is the energy conversion coefficient, $\left(0< \beta < 1\right)$; $P_U$ denotes the $U_M$ transmitted power; $t^{wpt}=\eta T$, $\eta$ is the \gls{eh} ratio,  $\left(0< \eta < 1\right)$; and $T$ is the protocol time. 
	
	\item In the secrecy offloading phase, denoted as $t_{\vartheta^*}^{off}$: the selected \glspl{Ed} concurrently offload their tasks via uplink \gls{noma} resulting in the composite signal obtained at $U_M$ as: $y_{U}^{MEC}=\sqrt{\frac{{{\rho }_{{{F}^{*}}U}}{{P}_{{{F}^{*}}U}}}{{{\overline{\mathbbmsl{L}}}_{{{F}^{*}}U}}}}\left( {{{\tilde{g}}}_{{{F}^{*}}U}}+{{\lambda }_{{{F}^{*}}U}} \right){{x}_{{{F}^{*}}U}}+\sqrt{\frac{{{\rho }_{{{N}^{*}}U}}{{P}_{{{N}^{*}}U}}}{{{\overline{\mathbbmsl{L}}}_{{{N}^{*}}U}}}}\left( {{{\tilde{g}}}_{{{N}^{*}}U}}+{{\lambda }_{{{N}^{*}}U}} \right){{x}_{{{N}^{*}}U}}+{{n}_{U}}$,
	where $\rho_{\vartheta^* U}$ is the power allocation ratio of $U_M$ $({{\rho }_{{{F}^{*}}U}}+{{\rho }_{{{N}^{*}}U}}=1$, $\text{ }{{\rho }_{{{F}^{*}}U}}>{{\rho }_{{{N}^{*}}U}})$; ${{P}_{{{\vartheta }^{*}}U}}=\frac{{{E}_{{{\vartheta }^{*}}U}}}{(1-\eta )T-t^{com}}$ is the power transmit of \glspl{Ed} to $U_M$; $x_{{{\vartheta }^{*}}U}$ is the transmitted signal of the optimal \glspl{Ed}; $n_U$ is subject to the \gls{awgn} distribution, with ${{{n}}_{U}}\sim \mathcal{C}\mathcal{N}\left( 0,{{\mathcal{N}}_{U}} \right)$. In the decoding processes, by following the \gls{noma} principles, $U_M$ initially tackles the $x_{F^*}$ signal by treating $x_{N^*}$ as interference. Thereafter, $U_M$ implements SIC to extract the preceding decoded signal. Hence, the expressions for the \glspl{sinr} at $U_M$ is expressed as \cite{NA23}:
	\begin{align} \label{sinr_U}
		\gamma _{{{F}^{*}}}^{U}=\frac{{{z}_{1}}{{a}^{2}}}{{{z}_{2}}{{b}^{2}}+{{z}_{3}}},  
		\gamma _{{{N}^{*}}}^{U}=\frac{{{z}_{2}}{{b}^{2}}}{{{\nu }_{1}}{{z}_{1}}{{a}^{2}}+{{z}_{4}}},
	\end{align}
	where ${{z}_{1}}=\frac{{{\gamma }_{{{F}^{*}}U}}}{{{\overline{\mathbbmsl{L}}}_{{{F}^{*}}U}}}$, ${{z}_{2}}=\frac{{{\gamma }_{{{N}^{*}}U}}}{{{\overline{\mathbbmsl{L}}}_{{{N}^{*}}U}}}$, ${{z}_{3}}={{z}_{1}}{\it\Omega}_{{{F}^{*}}U}^{2}+{{z}_{2}}{\it\Omega} _{{{N}^{*}}U}^{2}+1$,
	${{z}_{4}}={{\nu }_{1}}{{z}_{1}}{\it\Omega} _{{{F}^{*}}U}^{2}+{{z}_{2}}{\it\Omega} _{{{N}^{*}}U}^{2}+1$, ${{\gamma }_{{{F}^{*}}U}}=\frac{{{\rho }_{{{F}^{*}}U}}\beta {{\gamma }_{U}}{t^{wpt}}}{{{\overline{\mathbbmsl{L}}}_{{{F}^{*}}U}}\left[ \left( 1-\eta  \right)T-t^{com} \right]}$, ${{\gamma }_{{{N}^{*}}U}}=\frac{{{\rho }_{{{N}^{*}}U}}\beta {{\gamma }_{U}}{t^{wpt}}}{{{\overline{\mathbbmsl{L}}}_{{{N}^{*}}U}}\left[ \left( 1-\eta  \right)T-t^{com} \right]}$, ${{\gamma }_{U}}=\frac{{{P}_{U}}}{{{\mathcal{N}}_{U}}}$, $a={{\left| {{{\tilde{g}}}_{{{F}^{*}}U}} \right|}^{2}}$, $b={{\left| {{{\tilde{g}}}_{{{N}^{*}}U}} \right|}^{2}}$, $\nu_1$ denotes the residual interference caused by the \gls{isic} scenarios, with $\left( 0\le \nu_1 \le 1 \right)$, and $\nu_1=0$ is \gls{psic}. Similarly, the expressions of composite signals at $E$ is expressed as: ${{y}_{E}}=\sqrt{\frac{{{\rho }_{{{F}^{*}}E}}{{P}_{E}}}{\mathbbmsl{L}_{{{F}^{*}}E}}}\left( {{{\tilde{g}}}_{{{F}^{*}}E}}+{{\lambda }_{{{F}^{*}}E}} \right){{x}_{{{F}^{*}}E}}+\sqrt{\frac{{{\rho }_{{{N}^{*}}E}}{{P}_{E}}}{\mathbbmsl{L}_{{{N}^{*}}E}}}\left( {{{\tilde{g}}}_{{{N}^{*}}E}}+{{\lambda }_{{{N}^{*}}E}} \right){{x}_{{{N}^{*}}E}}+{{n}_{E}}$, where $\rho_{\vartheta^* E}$ is the power allocation of $E$, $({{\rho }_{{{F}^{*}}E}}+{{\rho }_{{{N}^{*}}E}}=1$, ${{\rho }_{{{F}^{*}}E}}>{{\rho }_{{{N}^{*}}E}})$, ${{n}_{E}}\sim \mathcal{C}\mathcal{N}\left( 0,{{\mathcal{N}}_{E}} \right)$. Supposing $E$ also conforms SIC principal, the \gls{sinr} expressions at $E$ is given as \cite{NA24}: 
	\begin{align} \label{sinr_E}
		\gamma _{{{F}^{*}}}^{E}=\frac{{{z}_{5}}x}{{{z}_{6}}y+{{z}_{7}}}, \gamma _{{{N}^{*}}}^{E}=\frac{{{z}_{6}}y}{{{z}_{8}}},
	\end{align}
	where ${{z}_{5}}=\frac{{{\rho }_{{{F}^{*}}E}}{{\gamma }_{E}}}{\overline{\mathbbmsl{L}}_{{{F}^{*}}E}}$, ${{z}_{6}}=\frac{{{\rho }_{{{N}^{*}}E}}{{\gamma }_{E}}}{\overline{\mathbbmsl{L}}_{{{N}^{*}}E}}$, ${{z}_{7}}={{z}_{5}}{{\it\Omega }_{{{F}^{*}}E}}+{{z}_{6}}{{\it\Omega }_{{{N}^{*}}E}}+1$, ${{z}_{8}}={{z}_{6}}{{\it\Omega }_{{{N}^{*}}E}}+1$, ${{\gamma }_{E}}=\frac{{{P}_{E}}}{{{\mathcal{N}}_{E}}}$, $x={{\left| {{{\tilde{g}}}_{{{F}^{*}}E}} \right|}^{2}}$, $y={{\left| {{{\tilde{g}}}_{{{N}^{*}}E}} \right|}^{2}}$.
	
	\item In the third phase, denoted as $t^{com}$: The offloaded tasks are calculated  at $U_M$. The duration necessary for the accomplishment of these computational processes with a number of task bits is expressed as: $t^{com}=\frac{\left( \mathcal{C}_{{{q}^{*}}}^{off}+\mathcal{C}_{{{j}^{*}}}^{off} \right)\varpi }{f_{MEC}^{}}$, where $\varpi$ is the number of CPU cycles requires to compute a single task bit and ${f_{MEC}^{}}$ is the \gls{mec}'s operating frequency at $U_M$.	
	
	\item In the fourth phase, denoted as $t^{down}$: Eventually, $U_M$ transfers the outcomes to the edge clusters through their corresponding optimal \gls{Ed}. Notably, due to the minimal energy depletion and brief latency associated with the data retrieval, the ${{t}^{down}}$ phase can be negligible \cite{NA23_2}.
\end{itemize}
\section{Secrecy Offloading Efficacy Analysis}\label{sec:pa}
\subsection{Preliminaries}
In the proposed system, the channel capacity of \gls{uav} $(U_M)$ and that of eavesdropper $(E)$ to detect the best remote/nearby \glspl{Ed} are given as follows, respectively: 
\begin{align} \label{capacity-U}
	C_{{{\vartheta }^{*}}}^{U}&=\left[ (1-\eta )T-{{t}^{com}} \right]W{{\log }_{2}}\left( 1+\gamma _{{{\vartheta }^{*}}}^{U} \right), \\ \label{capacity-E}
	C_{{{\vartheta }^{*}}}^{E}&=\left[ (1-\eta )T-{{t}^{com}} \right]W{{\log }_{2}}\left( 1+\gamma _{{{\vartheta }^{*}}}^{E} \right),
\end{align}
where $W$ is the system bandwidth. Thus, the secrecy capacity of \glspl{Ed} to $U_M$, with the interference of an passive eavesdropper is characterized as follows: 
\begin{align} \nonumber
	C_{{{\vartheta }^{*}}}^{S}&={{\left\lceil C_{{{\vartheta }^{*}}}^{U}-C_{{{\vartheta }^{*}}}^{E} \right\rceil }^{+}} \\ \label{capacity-Sec}
	& = \begin{cases}
		\left[ (1-\eta )T-{{t}^{com}} \right]W{{\log }_{2}}\left( \frac{1+\gamma _{{{\vartheta }^{*}}}^{U}}{1+\gamma _{{{\vartheta }^{*}}}^{E}} \right), & \gamma _{{{\vartheta }^{*}}}^{U}>\gamma _{{{\vartheta }^{*}}}^{E} \\
		0. & \gamma _{{{\vartheta }^{*}}}^{U}\le \gamma _{{{\vartheta }^{*}}}^{E}
	\end{cases}
\end{align}

\subsection{\Acrfull{sscp}}
This subsection delves into the derivation of the \gls{sscp} formulations, denoted by ${\mathcal{S}}^j_{s^{*}}$, where $j \in \left\{1,2\right\}$. The \gls{sscp} metric is determined as the likelihood that the duration of the offloaded tasks ${{t}^{off}_{{{\vartheta}^{*}}}}$ accomplishes within the prescribed latency threshold $T_{th}$, while the secrecy capacity ${{C}^{S}_{{{\vartheta}^{*}}}}$ functions above a predetermined rate ${{R}_{{{\vartheta}^{*}}}}$. Thus, the \gls{sscp} of the whole system is expressed as: 
\begin{align} \label{sim} \nonumber
	\mathcal{S}_{s^*}^j&=\Pr \left\{ t_{{{F}^{*}}}^{off}\le {{T}_{th}},t_{{{N}^{*}}}^{off}\le {{T}_{th}}, \right. \\  
	&\hspace{2.5cm}\left. C_{{{F}^{*}}}^{S}\ge {{R}_{{{F}^{*}}}},C_{{{N}^{*}}}^{S}\ge {{R}_{{{N}^{*}}}} \right\},
\end{align}
where $t_{{{\vartheta }^{*}}}^{off}=\frac{\mathcal{C}_{{{\vartheta }^{*}}}^{off}}{{{C}^U_{{{\vartheta }^{*}}}}}$, and ${{R}_{{{\vartheta}^{*}}}}=\frac{\mathcal{C}_{{{\vartheta}^{*}}}^{off}}{{{T}_{th}}}$. We assume that ${{T}_{th}}=(1-\eta )T-{t}^{com}$. Hence, the closed-formed formulation of \gls{sscp} is expressed as two cases in \textit{Lemma 1} and \textit{Lemma 2}.

\begin{lemma}
	The closed-form formulation for the \gls{sscp} of the whole system under Nakagami-$m$ fading channel in the case of \gls{isic} is derived as:
	\begin{align} \nonumber
		{{\mathcal{S}}^1_{{{s}^{*}}}}&={{\psi }_{1,c}}\sum\limits_{h}{\left( Q-1 \right)}\sum\limits_{t}{\left( K-1 \right)\Psi _{1}^{\left( {{\wp }_{o}} \right)}\Psi _{2}^{\left( \omega _{n} \right)}} \\ \nonumber &\times\left[ {{\delta }_{1}}-\delta _{2}^{\left( {{\Delta }_{4}} \right)} \right.-\delta _{3}^{\left( {{\Theta }_{1}} \right)}\left. \left( \delta _{4}^{\left( {{\Theta }_{1}} \right)}-\delta _{5}^{\left( {{\Delta }_{4}},{{\Theta }_{3}} \right)} \right) \right],
	\end{align}
	where ${{\psi }_{1,c}}=\frac{{{\pi }^{2}}\ddot{K}\ddot{Q}}{4NO\left( m-1 \right)!}{{\left( \frac{m}{{{\xi }_{{{N}^{*}}E}}} \right)}^{m}}$, $\Psi _{1}^{\left( {{\wp }_{o}} \right)}=\sum\limits_{o=1}^{O}$ ${\sqrt{1-\varphi _{o}^{2}}}\frac{\wp _{o}^{m-1+\bar{h}}{{\omega }_{o}}^{\frac{\wp _{o}^{2}m\left( h+1 \right)}{{{\xi }_{{{N}^{*}}U}}}-1}\left( \Delta _{3}^{\left( {{\wp }_{o}} \right)}-\Delta _{1}^{\left( {{\wp }_{o}} \right)} \right)}{{{\ln }^{2}}\left( {{\omega }_{o}} \right)}$, $\Psi _{2}^{\left( \omega _{n} \right)}=\sum\limits_{n=1}^{N}{\sqrt{1-\varphi _{n}^{2}}}{ \omega _{n}^{m-1+\bar{t}}}{{e}^{-\frac{m\left( t+1 \right)}{{{\xi }_{{{F}^{*}}U}}}\omega _{n}}}$, ${{\delta }_{1}}=\frac{\left( m-1 \right)!}{{\left( m/{{\xi }_{{{N}^{*}}E}} \right)}^{m}}$, $\delta _{2}^{\left( {{\Delta }_{4}} \right)}={{e}^{\frac{-m}{{{\xi }_{{{N}^{*}}E}}}\Delta _{4}^{\left( \omega _{n},{{\wp }_{o}} \right)}}}\sum\limits_{{{k}_{1}}=0}^{m-1}{\frac{\left( m-1 \right)!}{{{k}_{1}}!}\frac{{{\left( \Delta _{4}^{\left( \omega _{n},{{\wp }_{o}} \right)} \right)}^{{{k}_{1}}}}}{{{\left( m/{{\xi }_{{{N}^{*}}E}} \right)}^{m-{{k}_{1}}}}}}$, $\delta _{3}^{\left( {{\Theta }_{1}} \right)}={{e}^{\frac{-m}{{{\xi }_{{{F}^{*}}E}}}\Theta _{1}^{\left( \omega _{n},{{\wp }_{o}} \right)}}}\sum\limits_{s=0}^{m-1}{\frac{1}{s!}{{\left( \frac{m}{{{\xi }_{{{F}^{*}}E}}} \right)}^{s}}}\sum\limits_{{{k}_{2}}=0}^{s}{\left( \begin{matrix}
			s  \\
			{{k}_{2}}  \\
	\end{matrix} \right)}{{\left( \frac{{{z}_{6}}}{{{z}_{7}}} \right)}^{{{k}_{2}}}}$ $\times{{\left( \Theta _{1}^{\left( \omega _{n},{{\wp }_{o}} \right)} \right)}^{s}}$, $\delta _{4}^{\left( {{\Theta }_{1}} \right)}=\frac{\left( m-1+{{k}_{2}} \right)!}{{{\left( \Theta _{3}^{\left( \omega _{n},{{\wp }_{o}} \right)} \right)}^{m+{{k}_{2}}}}}$, $\delta _{5}^{\left( {{\Delta }_{4}},{{\Theta }_{3}} \right)}={{e}^{-\Delta _{4}^{\left( \omega _{n},{{\wp }_{o}} \right)}\Theta _{3}^{\left( \omega _{n},{{\wp }_{o}} \right)}}}\sum\limits_{k=0}^{m-1+{{k}_{2}}}\frac{\left( m-1+{{k}_{2}} \right)!}{k!}\frac{{{\left( \Delta _{4}^{\left( \omega _{n},{{\wp }_{o}} \right)} \right)}^{k}}}{{{\left( \Theta _{3}^{\left( \omega _{n},{{\wp }_{o}} \right)} \right)}^{m+{{k}_{2}}-k}}}$. The mentioned $\Delta_{1}$, $\Delta_{3}$, $\Delta_{4}$, $\Theta_{1}$, $\Theta_{3}$ are characterized as: 
	\begin{align}
		\Delta _{1}^{\left( {{\wp }_{o}} \right)}&=\sqrt{\frac{{{\zeta }_{{{F}^{*}}}}\left( {{z}_{2}}\wp _{o}^{2}+{{z}_{3}} \right)}{{{z}_{1}}}}, \\
		\Delta _{3}^{\left( {{\wp }_{o}} \right)}&=\sqrt{\frac{{{z}_{2}}\wp _{o}^{2}-{{z}_{4}}{{\zeta }_{{{N}^{*}}}}}{{{\nu }_{1}}{{z}_{1}}{{\zeta }_{{{N}^{*}}}}}}, \\
		\Delta _{4}^{\left( \omega _{n},{{\wp }_{o}} \right)}&=\left( 1+\frac{{{z}_{2}}\wp _{o}^{2}}{{{\nu }_{1}}{{z}_{1}}{{ \omega _{n}^2 }}+{{z}_{4}}}-{{\partial }_{{{N}^{*}}}} \right)\frac{{{z}_{8}}}{{{z}_{6}}{{\partial }_{{{N}^{*}}}}}, \\
		\Theta _{1}^{\left( \omega _{n},{{\wp }_{o}} \right)}&=\left( 1+\frac{{{z}_{1}}{{ \omega _{n}^2 }}}{{{z}_{2}}\wp _{o}^{2}+{{z}_{3}}}-{{\partial }_{{{F}^{*}}}} \right)\frac{{{z}_{7}}}{{{z}_{5}}{{\partial }_{{{F}^{*}}}}},  \\ 
		\Theta _{3}^{\left( {{\omega }_{n}},{{\wp }_{o}} \right)}&=\left( \frac{m}{{{\xi }_{{{N}^{*}}E}}}+\frac{m{{z}_{6}}\Theta _{1}^{\left( {{\omega }_{n}},{{\wp }_{o}} \right)}}{{{\xi }_{{{F}^{*}}E}}{{z}_{7}}} \right),
	\end{align}
	where ${{\wp }_{o}}=-\frac{1}{\ln \left( {{\omega }_{o}} \right)}$, ${{\omega }_{o}}=\frac{{{\varphi }_{o}}+1}{2}$, ${{\varphi }_{o}}=\cos \left( \frac{\pi \left( 2o-1 \right)}{2O} \right)$, $\omega _{n}=\frac{\left( {{\varphi }_{n}}+1 \right)\left( \Delta _{3}^{\left( {{\wp }_{o}} \right)}-\Delta _{1}^{\left( {{\wp }_{o}} \right)} \right)}{2}+\Delta _{1}^{\left( {{\wp }_{o}} \right)}$, ${{\varphi }_{n}}=\cos \left( \frac{\pi \left( 2n-1 \right)}{2N} \right)$, ${{\zeta }_{{{\vartheta}^{*}}}}={{2}^{\frac{\mathcal{C}_{{{\vartheta}^{*}}}^{off}}{W{{\left( {{T}_{th}} \right)}^{2}}}}}-1$, ${{\partial }_{{{\vartheta}^{*}}}}={{2}^{\frac{{{R}_{{{\vartheta}^{*}}}}}{W{{T}_{th}}}}}$, $N$ and $O$ are the complexity versus accuracy trade-off coefficient \cite{Judd}.
	
	\begin{proof}
		See Appendix A.
	\end{proof}
\end{lemma}

\begin{lemma}
	The closed-form formulation for the \gls{sscp} of the whole system under Nakagami-$m$ fading channel in the case of \gls{psic} is express as::
	\begin{align} \nonumber
		{{\mathcal{S}}^2_{{{s}^{*}}}}&={{\psi }_{2,c}}\sum\limits_{h}{\left( Q-1 \right)}\sum\limits_{t}{\left( K-1 \right)\Psi _{3}^{\left( {{\wp }_{o}} \right)}}\left[ \Psi _{4}^{\left( {{\Delta }_{1}} \right)} \right. \\ \nonumber
		& \times \left( {{\delta }_{1}}-\delta _{6}^{\left( {{\Delta }_{5}} \right)} \right)\left. -\Psi _{5}^{\left( {{\Delta }_{1}} \right)}\delta _{7}^{\left( {{\Theta }_{1}} \right)}\left( \delta _{8}^{\left( {{\Theta }_{3}} \right)}-\delta _{9}^{\left( {{\Delta }_{5}},{{\Theta }_{3}} \right)} \right) \right] ,
	\end{align}
	where ${{\psi }_{2,c}}=\frac{\pi {{e}^{-{{\Delta }_{2}}}}\ddot{K}\ddot{Q}}{2O\left( m-1 \right)!}{{\left( \frac{m}{{{\xi }_{{{N}^{*}}E}}} \right)}^{m}}$, $\Psi _{3}^{\left( {{\wp }_{o}} \right)}=\sum\limits_{o=1}^{O}{\sqrt{1-\varphi _{o}^{2}}}\wp _{o}^{m-1+\bar{h}}{{\omega }_{o}}^{\frac{m\left( h+1 \right)}{{{\xi }_{{{N}^{*}}U}}}-1}$, $\Psi _{4}^{\left( {{\Delta }_{1}} \right)}={{e}^{-\frac{m\left( t+1 \right)}{{{\xi }_{{{F}^{*}}U}}}\Delta _{1}^{\left( {{\wp }_{o}} \right)}}}\sum\limits_{{{k}_{3}}=0}^{m-1+\bar{t}}{\frac{\left( m-1+\bar{t} \right)!}{{{k}_{3}}!}\frac{{{\left( \Delta _{1}^{\left( {{\wp }_{o}} \right)} \right)}^{{{k}_{3}}}}}{{{\left( m\left( t+1 \right)/{{\xi }_{{{F}^{*}}U}} \right)}^{m+\bar{t}-{{k}_{3}}}}}}$, $\delta _{6}^{\left( {{\Delta }_{5}} \right)}={{e}^{\frac{-m}{{{\xi }_{{{N}^{*}}E}}}\Delta _{5}^{\left( {{\wp }_{o}} \right)}}}\sum\limits_{{{k}_{1}}=0}^{m-1}{\frac{\left( m-1 \right)!}{{{k}_{1}}!}\frac{{{\left( \Delta _{5}^{\left( {{\wp }_{o}} \right)} \right)}^{{{k}_{1}}}}}{{{\left( m/{{{\xi }_{{{N}^{*}}E}}} \right)}^{m-{{k}_{1}}}}}}$, $\Psi _{5}^{\left( {{\Delta }_{1}} \right)}=\frac{\pi {{e}^{-\Delta _{1}^{\left( {{\wp }_{o}} \right)}}}}{2N}\sum\limits_{n=1}^{N}{\sqrt{1-\varphi _{n}^{2}}}\wp _{n}^{m-1+\bar{t}}{{\omega }_{n}}^{\frac{m\left( t+1 \right)}{{{\xi }_{{{F}^{*}}U}}}-1}$, $\delta _{7}^{\left( {{\Theta }_{1}} \right)}={{e}^{\frac{-m}{{{\xi }_{{{F}^{*}}E}}}\Theta _{1}^{\left( {{\wp }_{n}},{{\wp }_{o}} \right)}}}\sum\limits_{s=0}^{m-1}{\frac{1}{s!}{{\left( \frac{m}{{{\xi }_{{{F}^{*}}E}}} \right)}^{s}}}\sum\limits_{{{k}_{2}}=0}^{s}{\left( \begin{matrix}
			s  \\
			{{k}_{2}}  \\
	\end{matrix} \right)}{{\left( \frac{{{z}_{6}}}{{{z}_{7}}} \right)}^{{{k}_{2}}}}{{\left( \Theta _{1}^{\left( {{\wp }_{n}},{{\wp }_{o}} \right)} \right)}^{s}}$, $\delta _{8}^{\left( {{\Theta }_{3}} \right)}=\frac{\left( m-1+{{k}_{2}} \right)!}{{{\left( \Theta _{3}^{\left( {{\wp }_{n}},{{\wp }_{o}} \right)} \right)}^{m+{{k}_{2}}}}}$, $\delta _{9}^{\left( {{\Delta }_{5}},{{\Theta }_{3}} \right)}={{e}^{-\Delta _{5}^{\left( {{\wp }_{o}} \right)}\Theta _{3}^{\left( {{\wp }_{n}},{{\wp }_{o}} \right)}}}$ $\times\sum\limits_{k=0}^{m-1+{{k}_{2}}}{\frac{\left( m-1+{{k}_{2}} \right)!}{k!}\frac{{{\left( \Delta _{5}^{\left( {{\wp }_{o}} \right)} \right)}^{k}}}{{{\left( \Theta _{3}^{\left( {{\wp }_{n}},{{\wp }_{o}} \right)} \right)}^{m+{{k}_{2}}-k}}}}$, ${{\Delta }_{2}}=\sqrt{\frac{{{z}_{4}}{{\zeta }_{{{N}^{*}}}}}{{{z}_{2}}}}$, $\Delta _{5}^{\left( {{\wp }_{o}} \right)}=\left( 1+\frac{{{z}_{2}}\wp _{o}^{2}}{{{z}_{4}}}-{{\partial }_{{{N}^{*}}}} \right)\frac{{{z}_{8}}}{{{z}_{6}}{{\partial }_{{{N}^{*}}}}}$. Noted in this \gls{psic} scenario, we assume that ${{\wp }_{o}}=-\ln \left( {{\omega }_{o}} \right)$, ${{\omega }_{o}}=\frac{\left( {{\varphi }_{o}}+1 \right){{e}^{-{{\Delta }_{2}}}}}{2}$, ${{\wp }_{n}}=-\ln \left( {{\omega }_{n}} \right)$, ${{\omega }_{n}}=\frac{\left( {{\varphi }_{n}}+1 \right){{e}^{-\Delta _{1}^{\left( {{\wp }_{o}} \right)}}}}{2}$.
	
	\begin{proof}
		See Appendix B.
	\end{proof}
\end{lemma}
\section{Numerical Results}\label{sec:nr}
This section presents the theoretical findings that verifies the offloading performance of the considered system. The parameter values across all the simulations is specified as follows \cite{NA24_EAI}: $(x_U,y_U) =(0,0)$, ${h_U} \in (50,300) $ (m),  $(x_F,y_F) =(-100,-100)$ (m), $(x_N,y_N) =(10,10)$ (m), $(x_E,y_E) =(80,80)$ (m), $\theta = 2$, ${\mu }^{LoS}=1.6$, ${\mu }^{NLoS}=23$, ${\tau }_{1}=0.1139$, ${{\tau }_{2}}=12.0870$,  $c = {3\times{10^8}}$, $f_c = 10^5$ (Hz), $W=10^2$ (MHz), $\gamma_U \in [0,40]$ (dB), $\eta \in (0,1)$, $\beta = 0.9$, $T=1$ (s), $\sigma_{F^*}=\sigma_{N^*}=0.5$, $l=10^2$ (bits), ${\rho_{\vartheta^*U}}={\rho_{\vartheta^*E}}=0.8$, $f_{MEC}=10^2$ (MHz), $\varpi=10^2$, $m=2$, $N=O=10^3$.

\begin{figure}[!htb]
	\begin{minipage}{0.48\textwidth}
		\centering
		\includegraphics[width=.65\linewidth]{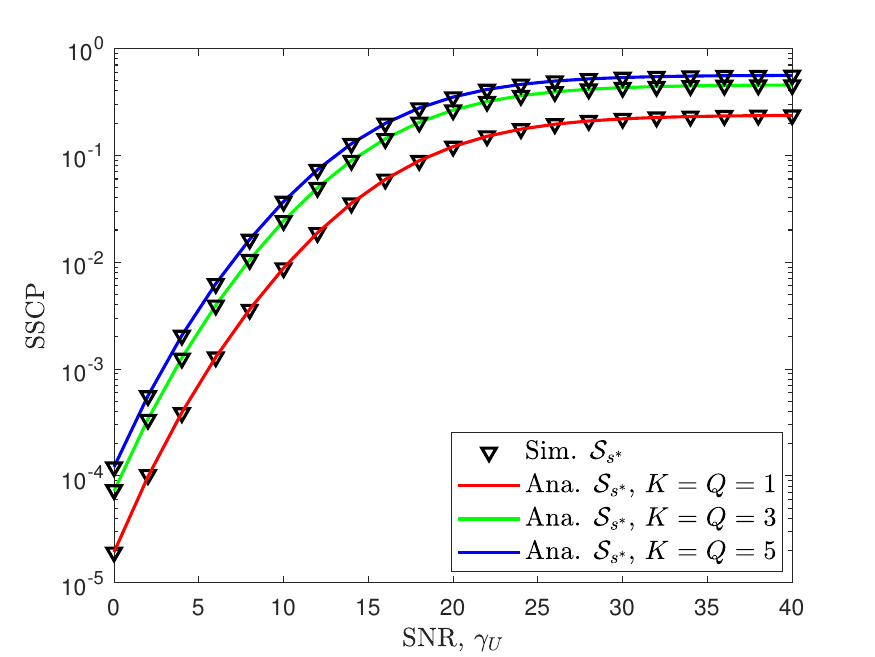}
		\caption{Effect of \gls{uav}'s average transmit \gls{snr}, $\gamma_U$ on \gls{sscp} of the whole system with a variety of \glspl{Ed}, with $\gamma_E=10$ (dB), $h_U=50$ (m), $\eta=0.7$, $\it\Omega$ $=3$, and $\nu_1=0.4$.} \label{fig:gU}
	\end{minipage}\hfill
	\begin{minipage}{0.48\textwidth}
		\centering
		\includegraphics[width=.65\linewidth]{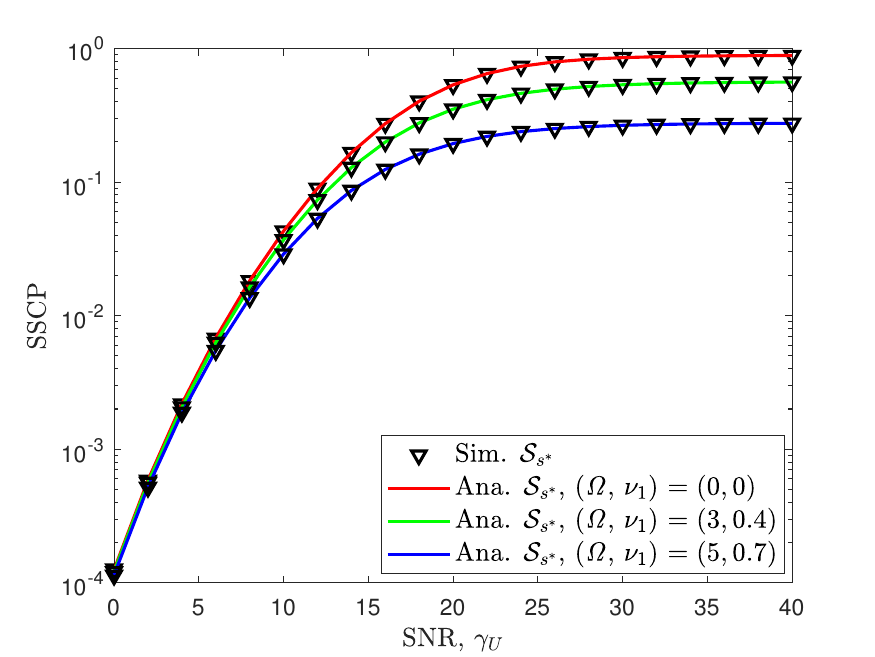}
		\caption{Effect of CSI parameters $\it\Omega$ and SIC parameters $\nu_1$ on \gls{sscp} of the whole system, with $K=Q=5$, $\gamma_E=10$ (dB), $h_U=50$ (m), and $\eta=0.7$.}\label{fig:pcsi-psic}
	\end{minipage}
\end{figure}
\begin{figure}[!htb]
	\begin{minipage}{0.48\textwidth}
		\centering
		\includegraphics[width=.65\linewidth]{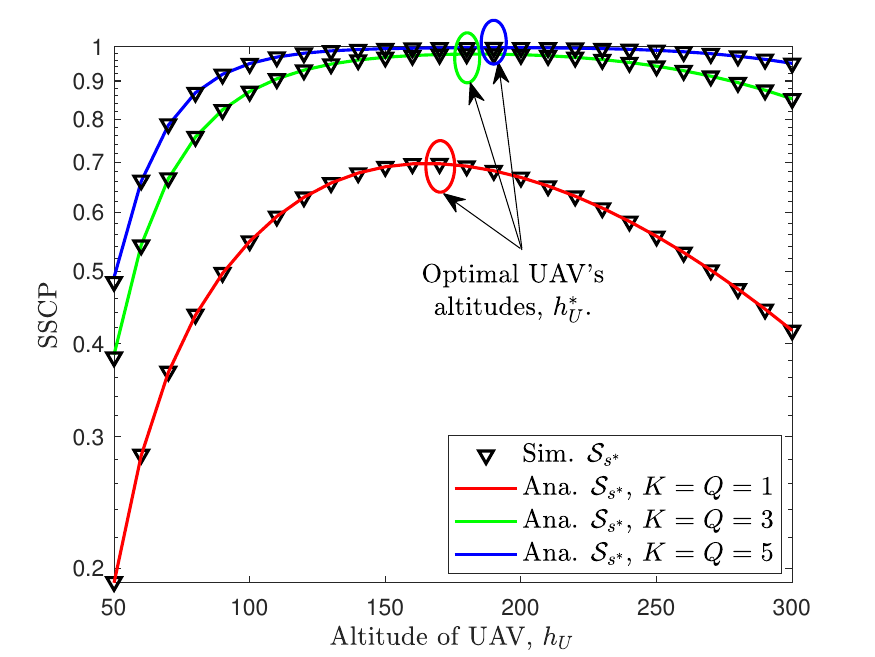}
		\caption{Effect of \gls{uav}'s altitude, $h_U$ on \gls{sscp} of the whole system with a variety of \glspl{Ed}, with $\gamma_U=30$ (dB), $\gamma_E=10$ (dB), $\eta=0.4$, ${\it\Omega}=3$, and $\nu_1=0.4$.} \label{fig:hU}
	\end{minipage}\hfill
	\begin{minipage}{0.48\textwidth}
		\centering
		\includegraphics[width=.65\linewidth]{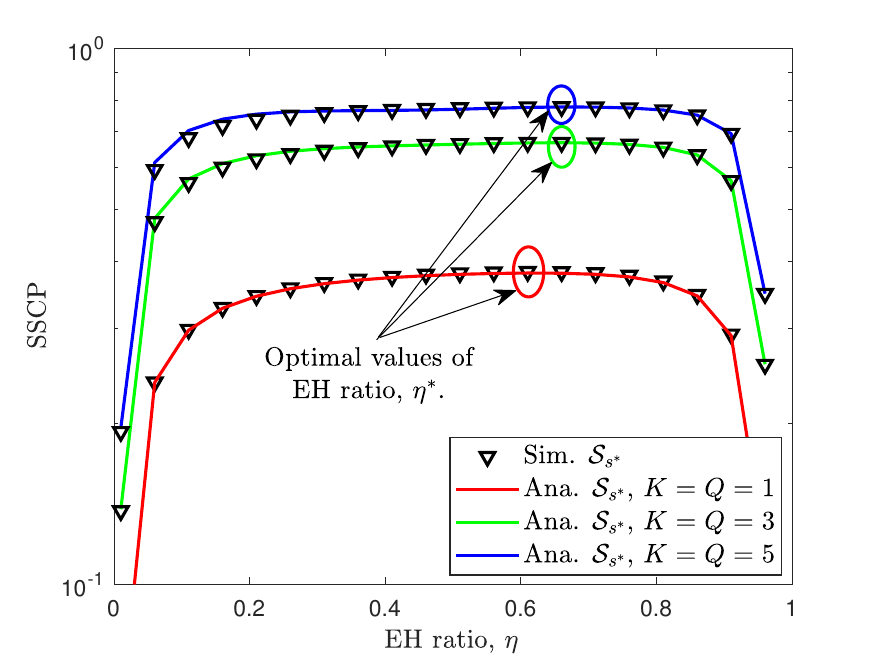}
		\caption{Effect of \gls{eh} ratio, $\eta$ on \gls{sscp} of the whole system with a variety of \glspl{Ed}, with $\gamma_U=38$ (dB), $\gamma_E=10$ (dB), $h_U=60$ (m), ${\it\Omega}=3$, and $\nu_1=0.4$.}\label{fig:eta}
	\end{minipage}
\end{figure}
\begin{figure}[!htb]
	\centering
	\includegraphics[width=.65\linewidth]{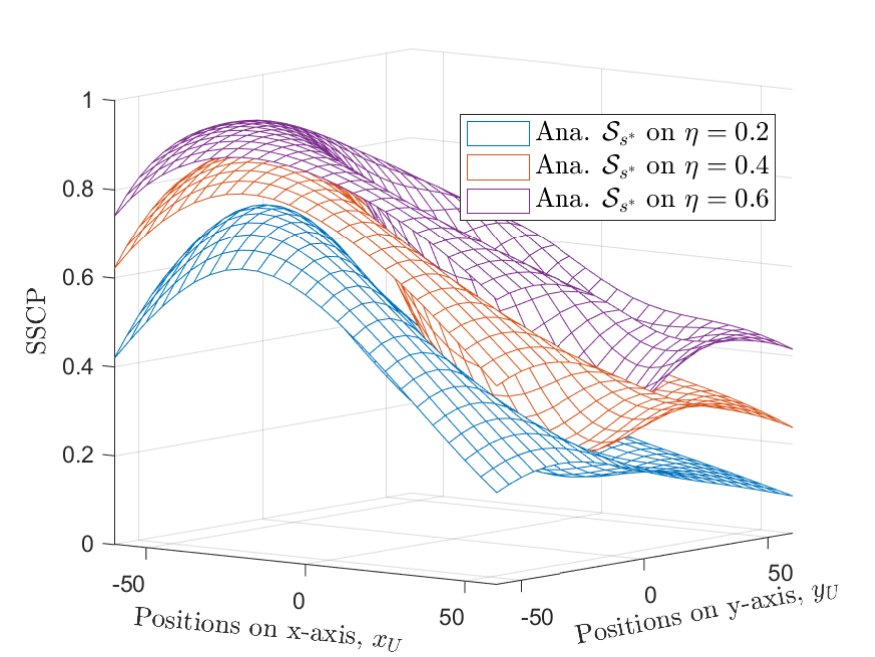}
	\caption{Effect of \gls{uav}'s position, $(x_U,y_U)$ on \gls{sscp} of the whole system with $K=Q=1$, $\gamma_U=34$ (dB), $\gamma_E=10$ (dB), $h_U=60$ (m), $\eta=0.4$, ${\it\Omega}=3$, and $\nu=0.4$.} \label{fig:3DxyU}
\end{figure}

As shown in Fig.~\ref{fig:gU}-\ref{fig:3DxyU}, the concurrence of the simulation (Sim.) and the theoretical study (Ana.) validates our system analysis. Fig.~\ref{fig:gU} depicts the effect of \gls{uav}'s transmission power $(\gamma_U)$ on \gls{sscp} of the whole system, with a multitude of \glspl{Ed} ($K$ and $Q$). The figure indicates that a concurrent rise in the device quantity enhances \gls{sscp}. This can be referred to an enlarged option of \glspl{Ed}, offering the \gls{uav} greater selection flexibility on secure data offloading. Additionally, an increased transmission source $\gamma_U$ boosts the \gls{sscp}, allowing \glspl{Ed} to posses plenty of power for effectively offloading

Fig.~\ref{fig:pcsi-psic} demonstrates the effect of CSI and SIC parameters on the \gls{sscp} of the whole system. The observation shows that the maximal system's secrecy offloading efficiency can be obtained under the state of pCSI-\gls{psic}. Nonetheless, the broadcast nature of channel conditions, associated with the inherent limitations in hardware, inevitably results in \gls{icsi}-\gls{isic} scenarios, thereby degrading the system performance.

Fig.~\ref{fig:hU} illustrates the effect of \gls{uav}'s altitudes on \gls{sscp} of the whole system. The result displays ideal points of \gls{uav}'s altitude $h_U^*$ that achieve the desirable \gls{sscp} values. This is attributed to the interchange between \gls{los} and \gls{nos} signal propagation of \gls{uav}-\glspl{Ed} reception. Elevating \gls{uav}'s altitude initially favors strong \gls{los} conditions, while an excessive altitude introduces greater path loss, consequently hindering the system secrecy efficacy.

Fig.~\ref{fig:eta} depicts the effect of \gls{eh} ratio $(\eta)$ on \gls{sscp} of the whole system. It identifies the optimal values of \gls{eh} ratio $\eta^*$ that maximize \gls{sscp}. As the initial increase in $\eta$ allows \glspl{Ed} to allocate more time for energy acquisition, hence enhancing the secrecy offloading operation. Alternatively, the exorbitant increase in $\eta$ diminishes the duration for offloading processes, thereby causing the reduction in \gls{sscp} efficacy.

To capitalize on the \gls{los} utility, the deployments of \gls{uav}'s positions is crucial for enhancing transmission link strength. Fig.~\ref{fig:3DxyU} depicts the effect of $(x_U,y_U)$ on \gls{sscp} of the whole system. The finding displays an optimal position ($x_U^*$, $y_U^*$) that optimizes \gls{sscp}. This finding is consistent with the intuition of the \gls{uav}'s strategic placements for optimal communication.

\section{Conclusion}\label{sec:conc}
In this paper, we studied an \gls{uav}-assisted \gls{noma}-integrated \gls{mec} with \gls{wpt} under Nakagami-$m$ fading channel in \gls{iot} networks. We introduced a quad-phased system protocol that ensures an efficient wireless charging and offloading operation. The \gls{sscp} closed-form formulations were expressed for the system's secrecy offloading efficacy assessments. Eventually, the theoretical findings were conducted with a variety of parameters that verifies the efficiency of the considered system.
\section*{Appendix A.}
By substituting the \eqref{sinr_U}, \eqref{sinr_E}, \eqref{capacity-U}, \eqref{capacity-Sec} into the defined formula of \eqref{sim}, ${\mathcal{S}}^1_{{{s}^{*}}}$ in the scenario of $\nu_1 >0$ can be expressed as: 
\begin{align} \label{eqA1} \nonumber
	  {\mathcal{S}}^1_{{{s}^{*}}}&=\int\limits_{0}^{\infty }{\int\limits_{\Delta _{1}^{\left( b \right)}}^{\Delta _{3}^{\left( b \right)}}{\int\limits_{0}^{\Delta _{4}^{\left( a,b \right)}}{{{F}_{x}}\left[ \left( 1+\frac{{{z}_{1}}{{a}^{2}}}{{{z}_{2}}{{b}^{2}}+{{z}_{3}}}-{{\partial }_{{{F}^{*}}}} \right)\frac{{{z}_{6}}y+{{z}_{7}}}{{{z}_{5}}{{\partial }_{{{F}^{*}}}}} \right]}}} \\ 
	  & \hspace{2.5cm}\times {{f}_{y}}\left( y \right){{f}_{a}}\left( a \right){{f}_{b}}\left( b \right)dydadb, 
\end{align}
where $\Delta _{1}^{\left( b \right)}=\sqrt{\frac{{{\zeta }_{{{F}^{*}}}}\left( {{z}_{2}}{{b}^{2}}+{{z}_{3}} \right)}{{{z}_{1}}}}$, $\Delta _{3}^{\left( b \right)}=\sqrt{\frac{{{z}_{2}}{{b}^{2}}-{{z}_{4}}{{\zeta }_{{{N}^{*}}}}}{{{\psi }_{1}}{{z}_{1}}{{\zeta }_{{{N}^{*}}}}}}$, $\Delta _{4}^{\left( a,b \right)}=\left( 1+\frac{{{z}_{2}}{{b}^{2}}}{{{\nu }_{1}}{{z}_{1}}{{a}^{2}}+{{z}_{4}}}-{{\partial }_{{{N}^{*}}}} \right)\frac{{{z}_{8}}}{{{z}_{6}}{{\partial }_{{{N}^{*}}}}}$. Thereby integrating \eqref{pdf-E} and \eqref{cdf-E} into \eqref{eqA1}, the first integral of ${\mathcal{S}}^1_{{{s}^{*}}}$ is derived as: 
\begin{align} \nonumber
	& {{I}_{1}}=\frac{1}{\left( m-1 \right)!}{{\left( \frac{m}{{{\xi }_{{{N}^{*}}E}}} \right)}^{m}}\left( \int\limits_{0}^{\Delta _{4}^{\left( a,b \right)}}{{{y}^{m-1}}{{e}^{-\frac{my}{{{\xi }_{{{N}^{*}}E}}}}}dy} \right. \\ \nonumber
	& -\int\limits_{0}^{\Delta _{4}^{\left( a,b \right)}}{{{y}^{m-1}}{{e}^{\frac{-my}{{{\xi }_{{{N}^{*}}E}}}}}{{e}^{\frac{-m}{{{\xi }_{{{F}^{*}}E}}}\left( 1+\frac{{{z}_{1}}{{a}^{2}}}{{{z}_{2}}{{b}^{2}}+{{z}_{3}}}-{{\partial }_{{{F}^{*}}}} \right)\frac{{{z}_{6}}y+{{z}_{7}}}{{{z}_{5}}{{\partial }_{{{F}^{*}}}}}}}}\sum\limits_{s=0}^{m-1}{\frac{1}{s!}} \\ \label{eqA2}
	& \left. \times {{\left( \frac{m}{{{\xi }_{{{F}^{*}}E}}} \right)}^{s}}{{\left[\left( 1+\frac{{{z}_{1}}{{a}^{2}}}{{{z}_{2}}{{b}^{2}}+{{z}_{3}}}-{{\partial }_{{{F}^{*}}}} \right)\frac{{{z}_{6}}y+{{z}_{7}}}{{{z}_{5}}{{\partial }_{{{F}^{*}}}}} \right]}^{s}}dy \right).
\end{align}
The first integral of \eqref{eqA2} can be solved by adopting the equation $\text{3}\text{.351}\text{.1}{{\text{.}}^{8}}$ in \cite{InT}. Thereafter, the equation $\text{1}\text{.111}$ of \cite{InT} is exploited in the second integral of \eqref{eqA2} for further analysis. Similarly applying the method in the $I_1$ first integral for the second integral, the completed expression of $I_1$ is achieved as:
\begin{align} \nonumber
	& {{I}_{1}}=\frac{1}{\left( m-1 \right)!}{{\left( \frac{m}{{{\xi }_{{{N}^{*}}E}}} \right)}^{m}}\left[ \frac{\left( m-1 \right)!}{{{\left( \frac{m}{{{\xi }_{{{N}^{*}}E}}} \right)}^{m}}}-{{e}^{\frac{-m}{{{\xi }_{{{N}^{*}}E}}}\Delta _{4}^{\left( a,b \right)}}} \right.\sum\limits_{{{k}_{1}}=0}^{m-1}{{}} \\ \nonumber
	& \frac{\left( m-1 \right)!}{{{k}_{1}}!}\frac{{{\left( \Delta _{4}^{\left( a,b \right)} \right)}^{{{k}_{1}}}}}{{{\left( \frac{m}{{{\xi }_{{{N}^{*}}E}}} \right)}^{m-{{k}_{1}}}}}-{{e}^{\frac{-m}{{{\xi }_{{{F}^{*}}E}}}\Theta _{1}^{\left( a,b \right)}}}\sum\limits_{s=0}^{m-1}{\frac{1}{s!}{{\left( \frac{m}{{{\xi }_{{{F}^{*}}E}}} \right)}^{s}}} \\ \nonumber
	& \times \sum\limits_{{{k}_{2}}=0}^{s}{\left( \begin{matrix}
			s  \\
			{{k}_{2}}  \\
		\end{matrix} \right)}{{\left( \frac{{{z}_{6}}}{{{z}_{7}}} \right)}^{{{k}_{2}}}}{{\left( \Theta _{1}^{\left( a,b \right)} \right)}^{s}}\left( \frac{\left( m-1+{{k}_{2}} \right)!}{{{\left( \Theta _{3}^{\left( a,b \right)} \right)}^{m+{{k}_{2}}}}} \right. \\ \label{eqA3}
	& \left. \left. -{{e}^{-\Delta _{4}^{\left( a,b \right)}\Theta _{3}^{\left( a,b \right)}}}\sum\limits_{k=0}^{m-1+{{k}_{2}}}{\frac{\left( m-1+{{k}_{2}} \right)!}{k!}\frac{{{\left( \Delta _{4}^{\left( a,b \right)} \right)}^{k}}}{{{\left( \Theta _{3}^{\left( a,b \right)} \right)}^{m+{{k}_{2}}-k}}}} \right) \right].
\end{align}
Combining with \eqref{pdf-U}, \eqref{cdf-U}, and \eqref{eqA3}, the second and final integral of ${\mathcal{S}}^1_{{{s}^{*}}}$ can be sequentially resolved with the aid of the Gaussian-Chebyshev Quadrature in \cite{Judd}. Thus, the closed-formed formulation of \gls{sscp} for the whole system in the case of \gls{isic} is derived as in \textit{Lemma 1}.

\section*{Appendix B.}
Similarly substituting as \eqref{eqA1}, ${\mathcal{S}}^2_{{{s}^{*}}}$ in the scenario of $\nu_1=0$ can be given as: 
\begin{align} \nonumber
	{\mathcal{S}}^2_{{{s}^{*}}}& =\int\limits_{{{\Delta }_{2}}}^{\infty }{\int\limits_{\Delta _{1}^{\left( b \right)}}^{\infty }{\int\limits_{0}^{\Delta _{5}^{\left( b \right)}}{{{F}_{x}}\left( \left( 1+\frac{{{z}_{1}}{{a}^{2}}}{{{z}_{2}}{{b}^{2}}+{{z}_{3}}}-{{\partial }_{{{F}^{*}}}} \right)\frac{{{z}_{6}}y+{{z}_{7}}}{{{z}_{5}}{{\partial }_{{{F}^{*}}}}} \right)}}} \\ \label{eqB1}
	& \hspace{2.5cm}\times {{f}_{y}}\left( y \right){{f}_{a}}\left( a \right){{f}_{b}}\left( b \right)dydadb,
\end{align}
where $\Delta _{5}^{\left( b \right)}=\left( 1+\frac{{{z}_{2}}{{b}^{2}}}{{{z}_{4}}}-{{\partial }_{{{N}^{*}}}} \right)\frac{{{z}_{8}}}{{{z}_{6}}{{\partial }_{{{N}^{*}}}}}$. The completed expression of the first integral in this case, denoted as $I_3$, is identically obtained as in \eqref{eqA3}. Hence, the expression of the second integral of ${\mathcal{S}}^2_{{{s}^{*}}}$ is obtained as: 
\begin{align} \label{eqB2}
	{{I}_{4}}=\int\limits_{\Delta _{1}^{\left( b \right)}}^{\infty }{{{f}_{a}}\left( a \right){{I}_{3}}da}.
\end{align}
By substituting \eqref{pdf-U} into \eqref{eqB2} and employing the distributive property, the $I_4$ integrals are achieved through the application of the equation $\text{3}\text{.351}\text{.2}{{\text{.}}^{11}}$ in \cite{InT} and the Quadrature method in \cite{Judd}. Eventually, the final integral is addressed by exploiting \eqref{pdf-U} along with the Quadrature method. Thus, the closed-formed formulation of \gls{sscp} for the whole system in the case of \gls{psic} is derived as in \textit{Lemma 2}.


\begin{thebibliography}{}
	\bibitem[1]{ZZh}Z. Zhi, “Research and Design of Industrial IoT Device Management System Based on 5G Communication and Big Data Technology,” 2nd ISPCEM, 2022.
	
	\bibitem[2]{JCo}J. Cook, Sabih ur Rehman, and M. Arif Khan, “Security and Privacy for Low Power IoT Devices on 5G and Beyond Networks: Challenges and Future Directions,” IEEE Access, vol.11, 2023.
	
	\bibitem[3]{NA22}A.-N. Nguyen, D.-B. Ha, V. N. Vo, V.-T. Truong, D.-T. Do, and C. So-In, “Performance Analysis of IoT Networks with UAV-assisted NOMA-based WPT-MEC,”  ICIIT, pp. 416-422, 2024.
	
	\bibitem[4]{SAh}S. Ahmed, Z. Subah, and M. Z. Ali, “Cryptographic Data Security for IoT Healthcare in 5G and Beyond Networks,” IEEE Sensors, 2022.
	
	\bibitem[5]{PSh}P. Shrivastava, Meeradevi, and M. R Mundada, “Survey on Congestion Control Approaches in 4G/5G Cellular Networks,” 4th I4C, 2022.
	
	\bibitem[6]{NWa}N. Wang, P. Wang, A. A.-Fanid, L. Jiao, and K. Zeng, “Physical Layer Security of 5G Wireless Networks for IoT: Challenges and Opportunities,” IEEE Int. of Things J., vol.6, no. 5, pp. 8169-8181, 2019.
	
	\bibitem[7]{RZh}R. Zhang, X. Pang, J. Tang, Y. Chen; N. Zhao, and X. Wang, “Joint Location and Transmit Power Optimization for NOMA-UAV Networks via Updating Decoding Order,” IEEE Wireless Communications Letters, vol. 10, no. 1, pp. 136-140, 2021.
	
	\bibitem[8]{NA24}A.-N. Nguyen, T.-S. Ngo, N.-A. Bui, P.-C. Le, and G.-H. Nguyen, “UAV-aided uplink NOMA based on MEC in IoT networks: Secrecy offloading and Optimization,” SaSeIoT, pp. 131-146, 2024.
	
	\bibitem[9]{NA22_2}A.-N. Nguyen, D.-B. Ha, V.-T. Truong, C. So-In, P. Aimtongkham, C. Sakunrasrisuay, and C. Punriboon, “On Secrecy Analysis of UAV-Enabled Relaying NOMA Systems with RF Energy Harvesting,” INISCOM, vol. 444, 2022.
	
	\bibitem[10]{HNG}G.-H. Nguyen, A.-N. Nguyen, H.-H. Le, T.-D. Do, “Energy Harvesting and Computation Offloading for UAV-Assisted MEC with NOMA in IoT Network,” J. Comm. and Int. Systems, vol. 3, pp. 381, 2023.
	
	\bibitem[11]{XZh}X. Zhang, J. Zhang, J. Xiong, L. Zhou, and J. Wei, “Energy-Efficient Multi-UAV-Enabled Multiaccess Edge Computing Incorporating NOMA,” IEEE Int. of Things J., vol. 7, no. 6, pp 5613-5627, 2021.
	
	\bibitem[12]{ISB}Ishan B., Neeraj K., Sudhanshu T., and Sudeep T., “Energy Consumption Minimization Scheme for NOMA-Based Mobile Edge Computation Networks Underlaying UAV,” IEEE Systems Journal, vol. 15, no. 4, pp 5724-5733, 2021.
	
	\bibitem[13]{NA23}A.-N. Nguyen, D.-B. Ha, T.-V. Truong, S. Sanguanpong, and C. So-In, “Secrecy Performance Analysis and Optimization for UAV-Relay-Enabled WPT and Cooperative NOMA MEC in IoT Networks,” IEEE Access, pp. 127800-127816, 2023.
	
	\bibitem[14]{KDo}K. Dogbe, and N. Hakem, “Analysis of full-duplex AF Relaying under Imperfect Channel State Information,” USNC-URSI, 2019.
	
	\bibitem[15]{NA23_2}A.-N. Nguyen, and N.-A. Bui, “Performance Analysis of IoT Mobile Edge Computing Networks Using a DF/AF UAV-Enabled Relay with Downlink NOMA,” ISIEA, 2023.
	
	\bibitem[16]{NA24_EAI}A.-N. Nguyen, T.-S. Ngo, N.-A. Bui, P.-C Le, M.-D Hoang, “Secrecy Offloading Analysis of NOMA-based UAV-aided MEC in IoT Networks with Imperfect CSI and SIC,” EAI End. Tran. Scal. Inf. Sys., vol. 11, 2024
	
	\bibitem[17]{InT}I. Gradshteyn and I. Ryzhik, Table of Integrals, Series, and Products. New York: Academic Press (Editor: A Jeffrey and D Zwillinger), 2014.
	
	\bibitem[18]{Judd}K. L. Judd, “Quadrature Methods,” University of Chicago’s Initiative for Computational Economics (ICE).
\end{thebibliography}
\end{document}